\documentclass[conference,10pt]{IEEEtran}
\IEEEoverridecommandlockouts
\usepackage{bookmark}
\usepackage{amsmath}
\usepackage{amsfonts}
\usepackage{bbding}
\usepackage{amssymb}
\usepackage{array}
\usepackage{subfigure}
\usepackage{graphicx}
\usepackage{subfigure}
\usepackage[named]{algo}
\usepackage{psfrag}
\usepackage{xfrac}
\usepackage{booktabs}
\usepackage{stfloats}
\usepackage{cases}
\usepackage[compress]{cite}
\usepackage{hyperref}
\usepackage{bm}
\usepackage{multirow}
\usepackage{amsthm}
\usepackage{stfloats}
\usepackage{setspace}
\usepackage{color}
\usepackage{algorithm}
\usepackage{algpseudocode}
\usepackage{amsmath}
\usepackage{cleveref}
\usepackage{makecell}
\usepackage{tabu}

\newcommand{\tabincell}[2]{\begin{tabular}{@{}#1@{}}#2\end{tabular}}
\makeatletter
\renewcommand{\citepunct}{,\penalty\@m\hskip.13emplus.1emminus.1em}
\renewcommand{\citedash}{\hbox{--}\penalty\@m}

\makeatother

\allowdisplaybreaks

\begin{document}

\title{Improving Learning Efficiency for Wireless Resource Allocation with Symmetric Prior}
\author{
 	\IEEEauthorblockN{Chengjian Sun, Jiajun Wu and Chenyang Yang}
	\IEEEauthorblockA{Beihang University, Beijing, China}
	Email:\{sunchengjian, jiajunwu, cyyang\}@buaa.edu.cn
	\thanks{The source code for reproducing the results of this article is available at {https://github.com/wjj19950101/Magrank-v1}.}
	\thanks{This work is supported by National Natural Science Foundation of China (NSFC) under Grant 61731002.}
}
\maketitle

\begin{abstract}
    Improving learning efficiency is paramount for learning resource allocation with deep neural networks (DNNs) in wireless communications over highly dynamic environments. Incorporating domain knowledge into learning is a promising way of dealing with this issue, which is an emerging topic in the wireless community.
    In this article, we first briefly summarize two classes of approaches to using domain knowledge: introducing mathematical models or prior knowledge to deep learning. Then, we consider a kind of symmetric prior, permutation equivariance, which widely exists in wireless tasks. To explain how such a generic prior is harnessed to improve learning efficiency, we resort to ranking, which jointly sorts the input and output of a DNN. We use power allocation among subcarriers, probabilistic content caching, and interference coordination to illustrate the improvement of learning efficiency by exploiting the property. From the case study, we find that the required training samples to achieve given system performance decreases with the number of subcarriers or contents, owing to an interesting phenomenon: ``sample hardening''. Simulation results show that the training samples, the free parameters in DNNs and the training time can be reduced dramatically by harnessing the prior knowledge. The samples required to train a DNN after ranking can be reduced by $15 \sim 2,400$ folds to achieve the same system performance as the counterpart without using prior.
\end{abstract}


\section{Introduction}
Machine learning has been envisioned as one of the most important features of 6G \cite{LYproc2020,RMPHY2020}, owing to its successful applications in a variety of complex tasks. In the past few years, deep learning has been widely applied in wireless communications \cite{RMPHY2020,Spawc2017,Samuel2019Lrn2Det,Zhou2019Beam,WJJ2020TCom}. The motivation is various, say improving spectral efficiency or user experience with low cost by providing real-time solutions for NP-hard optimization problems, handling the problems with inaccurate models or even without models, with well-trained deep neural networks (DNNs) \cite{Spawc2017,LYproc2020,Alessio2019Model}.

DNN is a powerful tool in expressing complex functions due to its multilayer structure and non-linear neurons, which has achieved great success in many fields such as computer vision and natural language processing \cite{Yann2015DL}.

Existing research results in wireless communications have demonstrated that learning-based solution is promising in improving the usage efficiency of radio resources in spatial, temporal, frequency, and power domain, as well as cache and computing resources. Yet training a DNN also consumes network resources, which is not negligible for learning at wireless edge. To enable intelligent wireless networks with affordable overall expense, learning efficiency is becoming another key performance indicator.

Learning efficiency can be reflected by generalization ability and training complexity. Generalization ability concerns with the system performance over a test set that differs from the training set for a given amount of training resources, i.e., data, computing and storage resources. Training complexity concerns with the amount of training resources, i.e., the numbers of training samples and trainable parameters as well as the training time,  required to achieve a desired system performance over the test set. These two metrics characterize the trade-off between the system performance in the inference phase and the resource consumption in the training phase.

Generalization ability has been extensively discussed in the literatures of machine learning and wireless communications. When used in static scenarios, DNNs can be trained offline, where training resources may be sufficient and hence are not a concern. However, as the development of wireless AI, machine learning has been introduced to mobile edges and expected to be applied in highly dynamic environment. For instance, radio resource allocation is often operated in time-varying fading channels, where training resources become the bottle-neck of the system performance. Whenever channels change, the DNN used for radio resource allocation has to be re-trained, and the training samples have to be re-gathered in a timely manner, otherwise the samples may be outdated \cite{Jie2021Online}. Hence, training has to complete in a short time with a small number of samples.

To reduce the training complexity required to achieve a desired performance, a rational choice is trading off flexibility with complexity by resorting to domain knowledge. Domain knowledge is quite general, including mathematical models for relations (say Shannon formula), structures of iterative algorithms, assumptions, and properties of task functions to be learned. To exploit the knowledge, two distinctive while compatible classes of approaches have been proposed in wireless communications. One class is to incorporate principled models, the other is to integrate prior knowledge, into the learning process.

Model-based deep learning takes the advantages of both data-driven and model-driven methods. Data-driven methods are flexible and hence are applicable to broad range of tasks, which however are with high training complexity and are non-interpretable. Model-driven methods strive to derive analytical expressions of resource allocation, by considering unique characteristics of a scenario without using any data. The resulting solutions are usually interpretable but only applicable to specific tasks.
By tailoring the structure of a DNN to a scenario of interest, model-based deep learning simplifies the task function to be learned. This can be achieved by only learning a part of a task that is hard to model, or only learning an operation or a block in a task with high complexity \cite{LYproc2020}. By customizing DNNs for specific tasks, model-based deep learning may need very few samples \cite{Shen2020LORM} or with low computational complexity for training \cite{Unfolding2020}.

Prior knowledge is the information of a task function available before training a DNN for the task. The knowledge for the input-output relation of a task can be leveraged for guiding the learning procedure, say by adding a regularization term in the loss function for training a DNN or by designing the DNN structure. By constructing DNNs to satisfy a desired property, the functions without the property are excluded from the function family, which reduces training complexity.
Two classical examples are convolutional neural networks and recurrent neural networks, which respectively exploit the knowledge of spatial and temporal translational invariance of a task \cite{Yann2015DL}. Another notable example is a class of symmetric DNNs \cite{Zaheer2017DeepSets}, which exploit a kind of prior knowledge broadly existed in wireless tasks \cite{Spawc2017,Zhou2019Beam, WJJ2020TCom, Alessio2019Model,Samuel2019Lrn2Det,GJ2020ICC,Eisen2020}: permutation equivariance (PE).

Permutation equivariance is a kind of symmetric property of multivariate functions. It has been harnessed by constructing various DNNs with special structures, say permutation equivariant neural network (PENN) \cite{Zaheer2017DeepSets} and graph neural networks \cite{Eisen2020}, for learning the policies with the property. These structures have been demonstrated capable of reducing the numbers of training samples and trainable parameters of DNNs \cite{GJ2020ICC} and generalizing well to the problem scales (say the number of users) \cite{Eisen2020}. As a kind of relational prior, this property can also be used for data representation and data argumentation.

In this article, we attempt to interpret how the learning efficiency of DNNs for permutation equivariant policies can be improved from the perspective of training complexity. In order to explain why many wireless tasks exhibit such a property, we first introduce the notions of object and the state of each object of a task, and figure out several kinds of objects commonly existed in resource allocation.
Given that interpreting DNNs is very hard, we explain the mechanism of reducing training complexity for tasks with symmetric property from the angle of data representation.
In particular, we jointly sort the input and output variables of a policy, called ranking. By taking ranking as a tool, we reveal that the reduction of training complexity comes from a fact that a task function with PE property can be learned in a small feature space.
In the cases where the states are scalars, we further find that the training samples required to achieve a given performance decrease dramatically with the number of objects, which comes from an interesting phenomenon: ``sample hardening'', i.e., the distribution of training samples is narrowed down after ranking.

The rest of this article is organized as follows. We first identify the PE property in wireless tasks. Then, we show how the property can be used to reduce sample complexity by ranking. Next, we provide case study to illustrate the potential of exploiting the property in reducing training complexity by comparing two approaches for exploiting the property, PENN and ranking, in the tasks of power allocation, caching and interference coordination, to the fully-connected DNN (FC-DNN) without using prior. Finally, we conclude the article and discuss the future works.

\section{A Generic prior knowledge in Wireless Tasks: Permutation Equivariance}
In this section, we show that permutation equivariant wireless tasks are widespread, ranging from physical layer to application layer. We first explain why many tasks have the PE property by figuring out their latent inputs, objects. Then, we provide the key components of a policy for such tasks with several concrete examples.

\subsection{What Wireless Tasks are Permutation Equivariant?}
Many problems in wireless communications aim to find a policy to achieve a desired performance, which yields a solution for every impacting parameter. The policy is usually a set of multivariate functions, where the input variables are  the impacting parameters and the output variables are the solution.
Representative tasks include (but not limited to) resource allocation in spatial-temporal-frequency-power domain \cite{Spawc2017,Zhou2019Beam,GJ2020ICC,Eisen2020}, transceiver design \cite{Samuel2019Lrn2Det}, uplink/downlink channel calibration\cite{Alessio2019Model}, and proactive caching \cite{WJJ2020TCom}.

\begin{figure}[!htbp]
  \centering
  \includegraphics[width=0.45\textwidth]{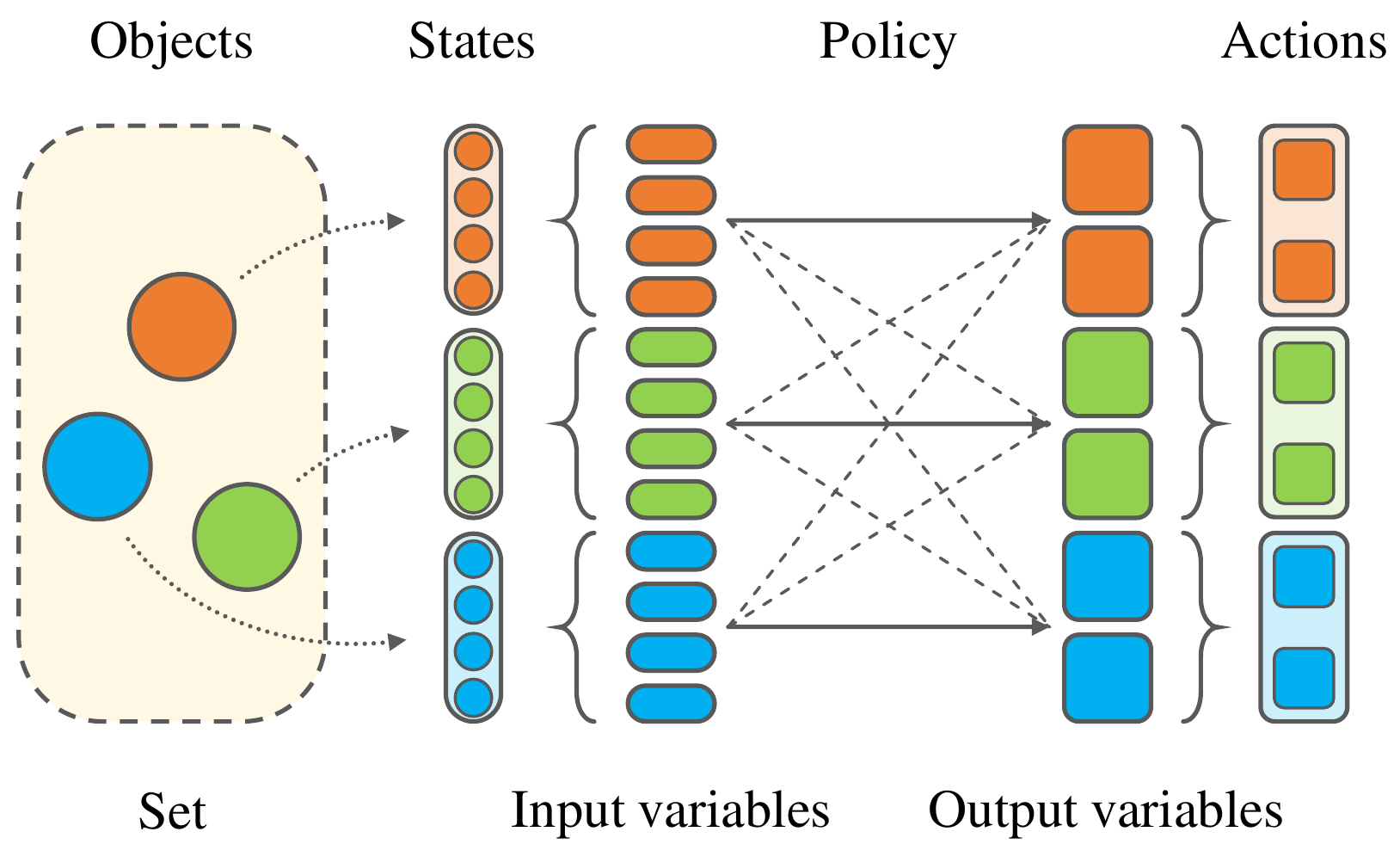}
  \caption{Relation between key components of a policy.}
  \label{Objects} \vspace{-2mm}
\end{figure}

These wireless policies are executed on a set of objects, e.g., users or contents, as illustrated in Fig.~\ref{Objects} with three objects. The input variables of a policy reflect the states of the objects relevant to the policy, say the channel gains of the users and the popularity of the contents (say files). The output variables of the policy reflect the actions taken on these objects, say the transmit powers allocated to the users and the caching probabilities for the files. The state or the action of each object itself can be multi-valued and expressed as a vector, noting that a scalar is a special case of a vector and a matrix can be expressed as a vector.
These objects compose a set (e.g., a user set), where each object is an element of the set.

A group of multivariate functions on a set must be permutation equivariant to the elements in the set \cite{Zaheer2017DeepSets}. In other words, the response of such a function is indifferent to the ordering of the elements.
If the state of every object can fully represent the useful information for the policy, then the policy must be permutation equivariant to the states. That is to say, a wireless policy for a set of objects will be permutation equivariant to the states, if the states are appropriately selected.

Despite that the PE property widely exists, it has been noticed in wireless communications only very recently \cite{Eisen2020,GJ2020ICC}, possibly due to the overlooking of the ``latent variables'': objects, over which the actions are taken.

\subsection{Several Examples of Resource Allocation}
Now we provide several typical tasks of resource allocation whose policies satisfy the PE property (called PE policies for short), and identify the objects and their states in each policy.

\subsubsection{Power Allocation among Subcarriers}
\begin{figure}[!htbp]
    \centering
    \includegraphics[width=0.45\textwidth]{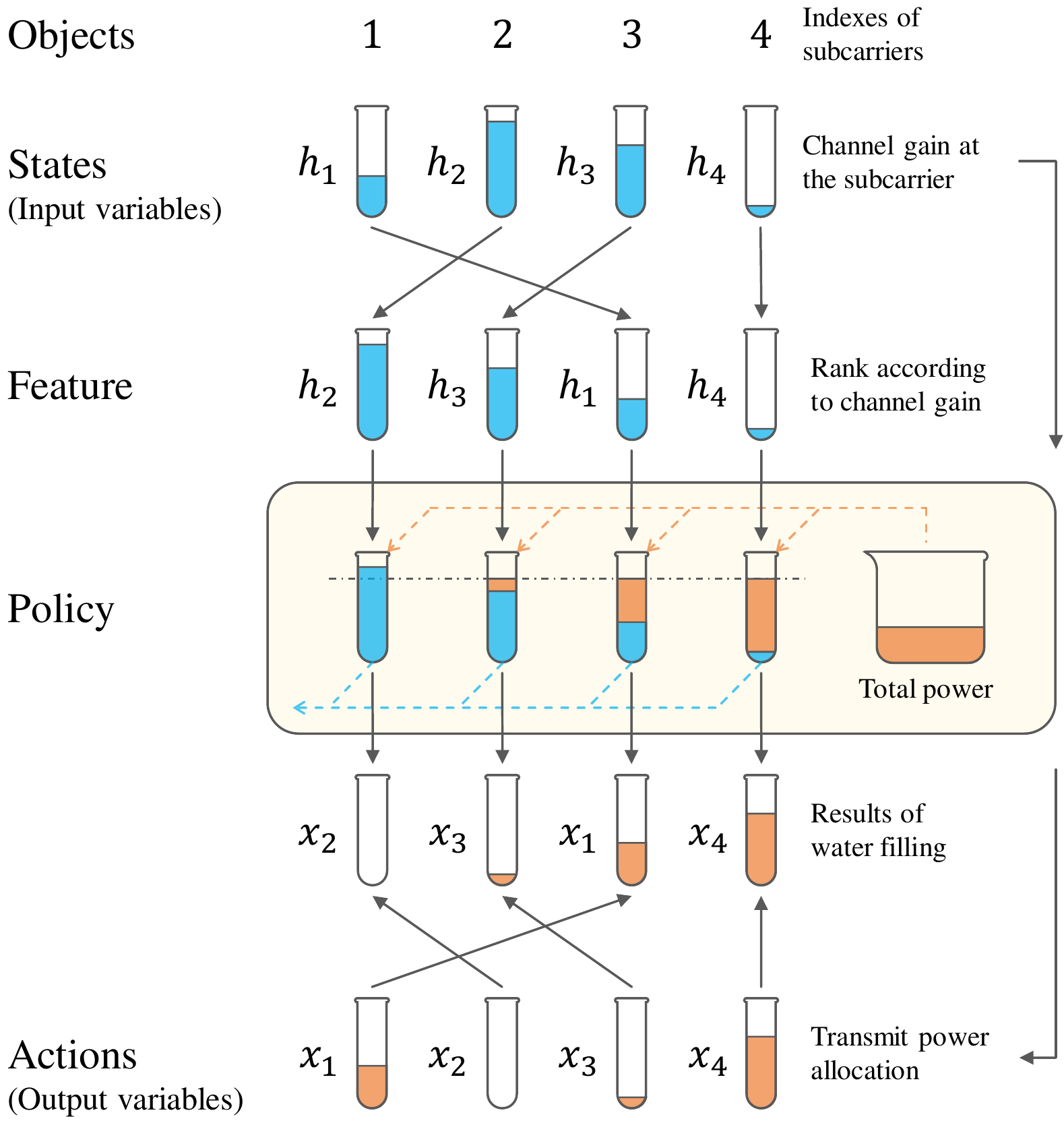}
    \caption{Permutation equivariance of a power allocation policy.}
    \label{PrmInv} \vspace{-1mm}
\end{figure}

We first consider a classical water-filling power allocation policy in a single user multi-subcarrier system. The  policy is a set of
multivariate functions, where the input and output variables are respectively the channel gains and transmit powers at the subcarriers. The policy is permutation equivariant since its response is indifferent to the ordering of the subcarriers.
A subcarrier is an object, the channel gain is the state of the object, and the transmit power allocated to the subcarrier is action. As illustrated in Fig.~\ref{PrmInv}, when the orders of the first, second and third objects change into the second, third and first, the orders of the input and output variables change in the same way whereas the policy remains unchanged.

\subsubsection{Interference Coordination}
\begin{figure*}[!htbp]
    \centering
    \includegraphics[width=0.95\textwidth]{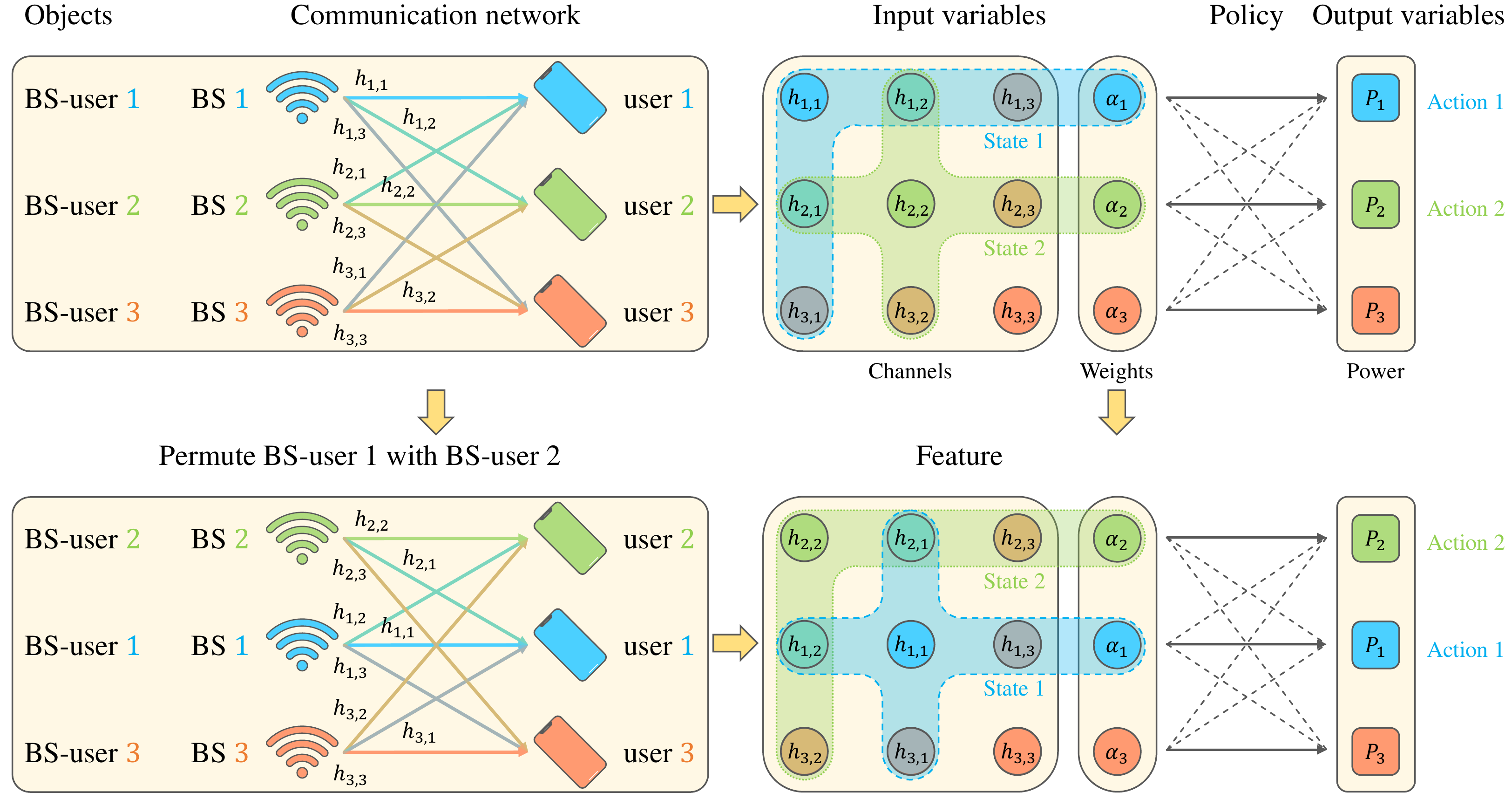}
    \caption{Permutation equivariance of an interference coordination policy.}
    \label{Channel} \vspace{-2mm}
\end{figure*}

Another typical example is interference coordination.
For easy exposition, consider a scenario in \cite{Spawc2017}, where each single-antenna base station (BS) serves a single-antenna user, and the interference is mitigated by controlling the transmit power of each BS to maximize a weighted sum rate. The policy is a set of
multivariate functions, where the input variables are the channels from all BSs to all users, and the output variables are the transmit powers at the BSs to their associated users. A BS and its associated user is an object, and the transmit power at the BS  is the action of the object. The action relies on the channels from the BS to all users, the channels from all BSs to the user, and the weight on the data rate, as shown in Fig.~\ref{Channel}. Hence, these channel gains and the weight are the state of the object. By defining the object and state in this way, the interference coordination policy is permutation equivariant to the BS-user pairs.
It is noteworthy that the policy to be learned for this task will not exhibit PE property if the state of an object does not contain the weight on the data rate when the weights are unequal. This illustrates the importance of appropriately defining the object and state in order to leverage the PE property inherent in a task.

\subsubsection{Caching at Wireless Edge}
In cache-enabled systems, the performance such as successful offloading probability (SOP) can be improved \cite{WJJ2020TCom} by optimizing probabilistic caching policy based on future file popularity. The policy is a set of functions, where the input variables are the popularity of all files, and the output variables are the caching probabilities for files. A file is an object, the future popularity of the file is the state,
and the future caching probability for the file is the action.

\section{How Training Complexity is Reduced?}
In this section, we first introduce a toy example to show the intuition of reducing training complexity for learning symmetric function. Then, we resort to ranking to explain how the training complexity is reduced for the PE policies.

To find a policy for a task with supervised learning, a DNN is trained with the samples each consisting of a realization of the random states and the corresponding
actions of the policy (i.e., the label). All possible training samples span an observation space. Without prior knowledge for the task, the policy can only be learned accurately by a non-structural DNN, i.e., fully-connected DNN (FC-DNN), from original data.
When prior knowledge is available, fewer training samples are required either by designing the DNN structure \cite{Yann2015DL,Zaheer2017DeepSets}, or by mapping the observation space where the data is gathered to the feature space where the samples are used.

\subsection{Reducing Training Complexity with Symmetric Prior}

A common practice in deep learning is to directly take the observed data as the feature for training. To understand why the training complexity for learning a task with symmetric property can be decreased by transforming observations to features, we provide a toy example.

Consider a task of learning an axial symmetric function, say quadratic function, with labeled training samples, where each sample contains a positive or negative input variable and the corresponding response of the function. The symmetric property of the function is the prior knowledge, with which the function can be determined by only given the observations of the function on the positive or negative real axis. Therefore, the sign of the input variable is uninformative for learning with the prior knowledge, and the absolute value of the input variable can be taken as the feature. After transforming each real-valued observation into a positive-valued feature, the training set is halved. Using the training samples each only containing positive input variable and the corresponding response, the function can be learnt with low complexity without sacrificing the learning performance.

\subsection{Ranking and Sample Hardening} \label{principle}
For easy visualization, let us consider a PE policy where the state of every object is a scalar. The policy for the objects is composed of multiple multivariate functions of a state vector and yields an action vector. The state vector of the policy consists of the states of all objects and spans the state space, which is the same as the observation space.

\begin{figure}[!htbp]
    \centering
    \includegraphics[width=0.45\textwidth]{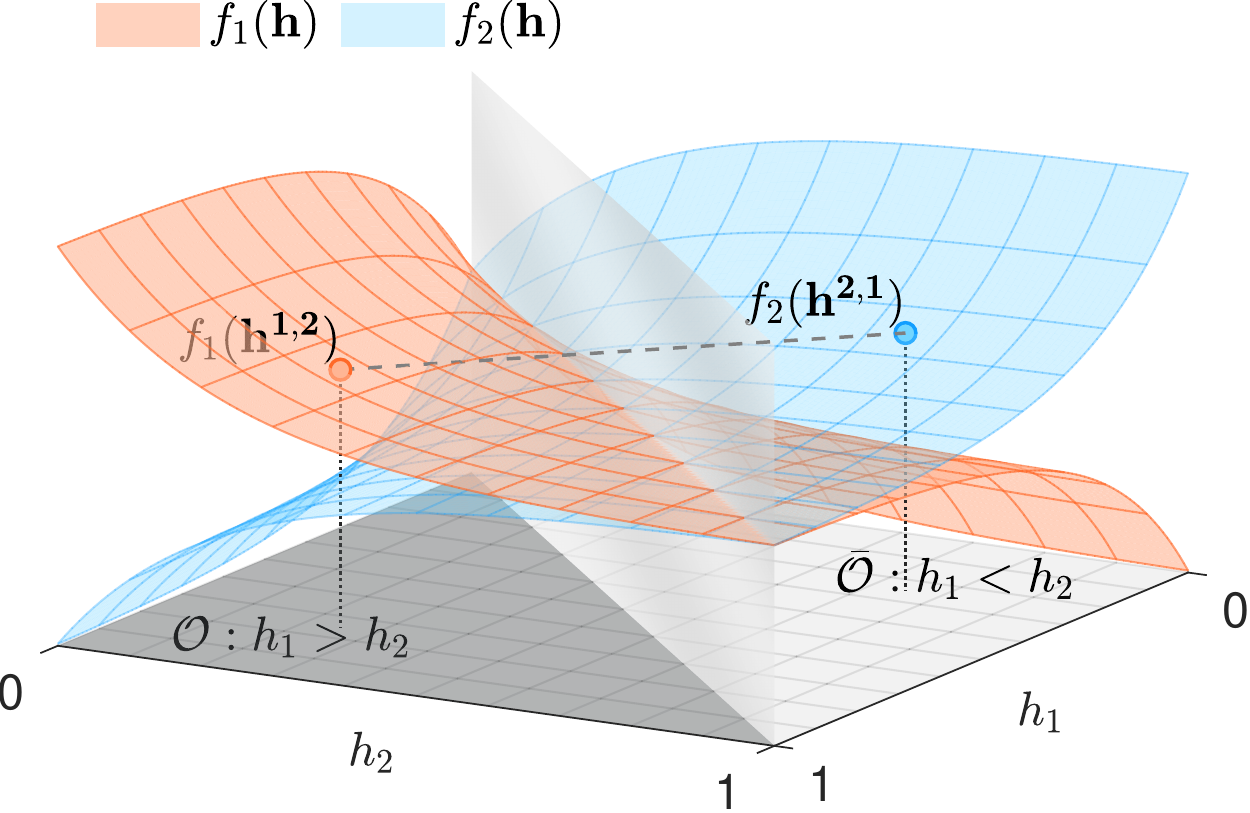}\vspace{-2mm}
    \caption{Illustration of mirror-symmetric functions. $h_1$ and $h_2$ are states of two objects, ${\bf h}^{1,2}=[h_1,h_2]$ and ${\bf h}^{2,1}=[h_2,h_1]$ are state vectors.}
    \label{Sym} \vspace{-2mm}
\end{figure}

For a PE policy, the order information of the objects implicitly embedded in the samples is useless for seeking the multivariate functions. To remove the order information, we can jointly sort the states and the labels in each sample according to a ranking rule that depends on the states. For instance, the states can be sorted in a descending order, and the actions are sorted accordingly. In this way, the state vectors with a same set of states arranged in different orders are mapped into a single feature vector, and so are the corresponding label vectors.

In Fig.~\ref{Sym}, we show two permutation equivariant multivariate functions for two objects, where each object has a single-valued state. The samples for the states satisfying $h_1 \!>\! h_2$, represented by a state vector ${\bf h}^{1,2}$, lie in the shadowed region $\mathcal{O}$.
Other samples, represented by state vector ${\bf h}^{2,1}$, fall in the non-shadowed region $\bar{\mathcal{O}}$, each can find its symmetric point ${\bf h}^{1,2}$ in $\mathcal{O}$ by swapping the two states.
Owing to the symmetric property, the responses of the functions in the non-shadowed region can be determined from the responses of the functions in the shadowed region.

This suggests that only a half of each symmetric function needs to be found in the halved observation space. Since the task function to be learned is simplified, the training complexity of the DNN can be reduced.

Moreover, the distribution of training samples is changed by ranking, since the state vectors only take values in the shrunken region. In particular, the distribution of the first element (and also the second element) of the two state vectors is narrower than the probability distribution of the non-ranked states. According to the theory of order statistic \cite{OSbook}, the variance of an element in any given position of a random vector with large number of elements approaches zero if the elements are ranked in {ascending or descending} order. This implies that ranking leads to ``sample hardening'' for a PE policy with scalar state for every object. As a consequence, the required training samples for learning a PE policy can be decreased dramatically when the number of objects is large.

When the state is multi-valued (say for the interference coordination policy), ranking is still able to reduce the sample complexity. However, finding a good metric that maps the multiple values in the state to a real number for ranking is challenging, which requires domain knowledge.

\section{Case Study}
In this section, we illustrate the gain in reducing training complexity from harnessing the PE property, by learning to optimize the three policies in section II with DNNs trained by supervision.

The training complexity includes the number of free parameters in each DNN, as well as the minimal number of samples and the computing time required for training each DNN to achieve (almost) the same system performance. All results are obtained with TensorFlow 1.14.0 on a computer with Intel\textsuperscript{\textregistered}-Core\textsuperscript{\texttrademark}-i7-8700K CPU and a single NVIDIA\textsuperscript{\textregistered}-GeForce-GTX\textsuperscript{\texttrademark}-1080-Ti GPU.

\subsection{{System Setups and Data Generation}}
The power allocation policy is optimized to maximize the sum rate of all subcarriers under the total transmit power constraint. The separation of subcarriers is $1$~MHz, and the signal-to-noise ratio is $10$~dB. {The input of the DNNs consists of the channel gains of all subcarriers, whose samples are generated from Rayleigh distribution. The output of the DNNs consists of the powers allocated to all subcarriers, whose labels are obtained from the classical water-filling algorithm.}

The interference coordination policy is optimized to maximize the sum rate of all users under the maximal transmit power constraint. The system setup is the same as the Gaussian interference channel case with equal weights in \cite{Spawc2017}. {The input of the DNNs consists of the channel gains among all BS-user pairs, whose samples are generated from Rayleigh distribution. The output of the DNNs consists of the transmit power of all BSs, whose labels are obtained from the weighted minimum mean square error (WMMSE) algorithm as in \cite{Spawc2017}.}

The caching policy is optimized to maximize the SOP, i.e., the probability that the data rate exceeds a threshold for a requested file cached at BSs. The system setup is the same as in \cite{WJJ2020TCom} where $10\%$ files are cached at each BS, except that here we consider homogeneous network. {The input of the DNNs consists of the future popularities of all files, whose samples are generated from Zipf distribution with the skewness parameter as $0.6$. The output of the DNNs consists of the caching probabilities of all files, whose labels are obtained from the water-filling algorithm in \cite{WJJ2020TCom}.}

\subsection{Performance Comparison}
We compare the performance of each policy learned by the following three DNNs.
\begin{itemize}
	\item ``W/o-Prior": FC-DNN trained by the samples without ranking, which does not exploit any prior knowledge.
	\item  ``Rank":  FC-DNN trained by the samples with ranking, which exploits the PE prior by data representation.
	\item  ``PENN":  PENN trained by the samples without ranking, which exploits the PE prior by DNN structure.
\end{itemize}

Without ranking, each original sample used for training or testing consists of the input of the DNN and the corresponding label. Each sample for the DNN to learn the power allocation policy (or the caching policy) consists of a channel vector (or a popularity vector) of the states and a column vector of the actions. Each sample for the DNN to learn the interference coordination policy can be expressed as a matrix, consisting of a channel matrix of the states and a column vector of the actions.

With ranking, each sample used for training or testing the DNNs to learn the power allocation and the caching policies is obtained by sorting the input and output vectors of the original sample in a descending order according to the magnitudes of the input of the DNN. Each sample used for the DNN to learn the interference coordination policy is obtained by permuting both column and row of the matrix in the same manner (e.g., permute the first and second columns and permute the first and second rows at the same time, as shown in Fig.~\ref{Channel}). Given that each state is composed of multiple channel gains, we sort each sample according to the channel gains between BSs and their associated users (i.e., the diagonal values of the channel matrix) in descending order, for an illustration.

{The generated data set is divided into training set, validation set and test set. The size of the training set is not fixed in order to evaluate the sample complexity. The size of the validation set is the minimal integer no smaller than $10\%$ of the size of the training set, and the test set contains $1,000$ samples.}
The DNNs are trained by minimizing the empirical mean square error between the labels and the outputs on the training set. {The training samples are divided into batches. Each batch is used to compute the gradient for updating the trainable parameters in DNNs in each iteration. To exploit the limited samples, the training set is repeatedly used for multiple times, counted in epochs.}
The hyper-parameters are fine-tuned on the validation set and are shown in Table~\ref{Hypara}. {Here, $[*_1, *_2, \cdots]$ denotes there are $*_1$ neurons in the 1st hidden layer, $*_2$ in the 2nd and so on, and the activation functions are the non-linear functions used in the neurons.
}

\begin{table*}[!htbp]
\footnotesize
\centering
\caption{Fine-tuned Hyper-parameters}
\label{Hypara}
\begin{tabular}{c|c|c|c|c|c|c|c|c|c|c}
\hline
\multicolumn{2}{c|} {Case }& \multicolumn{3}{c|}{Power allocation} & \multicolumn{3}{c|}{Caching} & \multicolumn{3}{c}{Interference coordination}\\
\hline
\multicolumn{2}{c|}{ \tabincell{c}{Number of objects}} &$10$ & $20$ & $30$ & $10$& $20$ & $30$ & $10$ & $20$ & $30$\\
\hline
\multirow{3}*{  \tabincell{c}{Number of neurons \\ in hidden layers} } & W/o-Prior  & [100] & [100] & [100]   & [50] & [90] & [120] & \multicolumn{3}{c}{ \tabincell{c}{ [200, 80, 80] \cite{Spawc2017}  } }  \\
\cline{2-11}
& Rank & [10]  & [5] & [5]  & [20] & [5] & [4]  &\multicolumn{3}{c}{ \tabincell{c}{  [150, 50, 50]  } }  \\
\cline{2-11}
& PENN & [100] & [200]  & [300] & [100] & [400] & [300] & [100, 100]  & [200, 200] & [300, 300]  \\
\hline
\multirow{3}*{  \tabincell{c}{Activation function \\ of the hidden layers} } & W/o-Prior  & \multicolumn{3}{c|}{ \multirow{3}*{  \texttt{ReLU} } } & \multicolumn{3}{c|}{ \multirow{3}*{ \texttt{ ReLU} } } & \multicolumn{3}{c}{ \multirow{3}*{\texttt{ReLU} \cite{Spawc2017} } }\\
\cline{2-2}
& Rank & \multicolumn{3}{c|}{}  & \multicolumn{3}{c|}{}\\
\cline{2-2}
& PENN & \multicolumn{3}{c|}{} & \multicolumn{3}{c|}{}  \\
\hline
\multirow{3}*{  \tabincell{c}{Activation function \\ of the output layer} } & W/o-Prior  &  \multicolumn{3}{c|}{ \multirow{3}*{ \texttt{Softplus}}}  &  \multicolumn{3}{c|}{ \multirow{3}*{ \texttt{Sigmoid} }}& \multicolumn{3}{c}{ \multirow{3}*{\texttt{ReLU6/6} \cite{Spawc2017} }}\\
\cline{2-2}
& Rank & \multicolumn{3}{c|}{ }&  \multicolumn{3}{c|}{ }\\
\cline{2-2}
& PENN &\multicolumn{3}{c|}{ } &  \multicolumn{3}{c|}{ }\\
\hline
\multirow{3}*{ \tabincell{c}{ Learning algorithm} } & W/o-Prior  &  \multicolumn{3}{c|}{ \multirow{2}*{ Adam (Initial: 0.1)} }  & \multicolumn{3}{c|}{ \multirow{3}*{ Adam (Initial: 0.01)} }  &  \multicolumn{3}{c}{ \multirow{2}*{RMSProp (Initial: 0.001) \cite{Spawc2017}  } }\\
\cline{2-2}
& Rank &   \multicolumn{3}{c|} {} & \multicolumn{3}{c|} {} \\
\cline{2-5}\cline{9-11}
& PENN & \multicolumn{3}{c|}{Adam (Initial: 0.001)}  & \multicolumn{3}{c|} {}  &  \multicolumn{3}{c}{Adam (Initial: 0.01)}   \\
\hline
\multirow{3}*{ \tabincell{c}{ Batch size} } & W/o-Prior  & 32 & 32 & 32 & 128 & 128 & 128 &  \multicolumn{3}{c}{ \multirow{2}* {1,000 \cite{Spawc2017}}} \\
\cline{2-8}
& Rank & 20  & 6  & 3  & 15 & 8 & 5     \\
\cline{2-11}
& PENN & 20  & 50  & 150  & 15 & 50 & 100  & 120 & 800 & 1,000  \\
\hline
\multirow{3}*{ \tabincell{c}{ Epochs} } & W/o-Prior  & 3,000 & 3,000 & 3,000 & 3,000 &3,000 & 3,000 &  300 &300 & 300  \\
\cline{2-11}
& Rank & 3,000 & 3,000 & 3,000 & 10,000 & 10,000 & 10,000   & 500 & 500 & 500  \\
\cline{2-11}
& PENN & 10,000  & 10,000  & 15,000  & 10,000 & 10,000 & 15,000  & 500 & 3,000 & 8,000  \\
\hline
\end{tabular}
\end{table*}

%\subsection{Performance Comparison}
In what follows, we evaluate the system performance achieved by each resource allocation policy learned by each well-trained DNN on the test set and the corresponding training complexity.

For power allocation or interference coordination, the system performance is the ratio of the sum rate achieved by the learning methods to the optimal solution or the solution obtained by the WMMSE algorithm. For probabilistic caching, the system performance is the ratio of the SOP achieved by the learning methods to the solution obtained from the water-filling algorithm in \cite{WJJ2020TCom}.

Since only a few training samples are required for the power allocation or caching policy, we train the DNNs more than once to show the impact of the random training set.
For each time of training, the training and validation samples are randomly generated, while the test set is fixed. The sum rate of the power allocation policy and the SOP of the caching policy are obtained by selecting the second worst testing results from 10 well-trained DNNs, hence the results are with the confidence level of 90\%. Because in the case of 30 BS-user pairs the computing time for training a PENN for interference coordination is too long, and the sum rate achieved by a PENN may be low (say 0.6 or 0.7), the performance of the interference coordination policy is obtained only from three well-trained DNNs by selecting the best testing results.
Even though, the sum rate achieved by ``PENN" is still lower than ``Rank'' and ``W/o-Prior'', which cannot be improved by further increasing training samples and free parameters in the DNN according to our results.

\begin{table*}[!htbp]
\footnotesize
\centering
\caption{Comparison of System Performance and Training Complexity.}
\label{PerfComp}
\begin{tabular}{c|c|c|c|c|c|c|c|c|c|c}
\hline
\multicolumn{2}{c|} {Case }& \multicolumn{3}{c|}{Power allocation} & \multicolumn{3}{c|}{Caching} & \multicolumn{3}{c}{Interference coordination}  \\
\hline
\multicolumn{2}{c|}{ \tabincell{c}{Number of objects}} &$10$ & $20$ & $30$ & $10$& $20$ & $30$ & $10$ & $20$ & $30$  \\
\hline
\multirow{3}*{  \tabincell{c}{System \\ performance} } & W/o-Prior  &  0.9948 & 0.9992 & 0.9999  & 0.9904 & 0.9909 & 0.9912 & 0.9795 & 0.9063 & 0.8562 \\
\cline{2-11}
& Rank & 0.9943 & 0.9951 &  0.9999 & 0.9900 & 0.9903 & 0.9915 & 0.9784 & 0.9042 & 0.8560\\
\cline{2-11}
& PENN & 0.9926 &  {\bf 0.9860} & {\bf 0.9770} & 0.9925 & 0.9903 & {\bf 0.9739} & {\bf 0.9003} & {\bf 0.8571} & {\bf 0.8418}  \\
\hline
\multirow{3}*{  \tabincell{c}{Training samples} } & W/o-Prior  &300 & 900  & 1,350  & 5,000 & 9,000 & 12,000 & 500,000& 1,000,000 & 1,000,000  \\
\cline{2-11}
& Rank & 20  & 6  & 3 & 15 & 8 & 5  &   10,000 & 3,000 & 2,000\\
\cline{2-11}
& PENN & 20 & 50 & 150 & 15 & 50 & 100 & 120 & 800 &  4,000 \\
\hline
\multirow{3}*{ \tabincell{c}{Free parameters} } & W/o-Prior  & 2,110 & 4,120 & 6,130 & 1,060 & 3,710 & 7,350& 25,570& 28,380 & 31,190 \\
\cline{2-11}
& Rank & 220 & 225 & 335 & 430 & 225 & 274 & 12,260& 14,070 & 16,080 \\
\cline{2-11}
& PENN & 51 & 51 &  51 & 51 & 101 & 51  & 480 & 480 & 480 \\
\hline
\multirow{3}*{ \tabincell{c}{ Training time \\ in seconds} } & W/o-Prior  & 21.64 & 61.2 & 89.01 & 90.96 & 165.61 & 225.78 & 574.10 & 4,198.18 & 9,844.96 \\
\cline{2-11}
& Rank & 5.45 & 5.51 & 5.52 & 18.14 & 18.19 & 18.48 & 16.88 & 16.52 & 24.02 \\
\cline{2-11}
& PENN & {31.74} & 32.05 &  50.32 & 31.70 & 32.02 & 50.29  & 6.28 & 219.16 & 6,737.43 \\
\hline
\end{tabular}
\end{table*}

The performance is provided in Table~\ref{PerfComp}. We can observe that ``Rank" and ``PENN" require much less training samples and free parameters to achieve similar performance to ``W/o-Prior". Moreover, the training samples required by ``Rank" decreases with the number of objects. In particular, only three or five training samples are required for the power allocation policy or caching policy with 30 objects, with a compression rate of $1,350/3=450$ or $12,000/5=2,400$ with respect to ``W/o-Prior". In fact, our result shows that only one training sample is required for the power allocation policy to achieve the system performance of 0.99 when there are 35 subcarriers! Such a surprising result comes from the ``sample hardening'' phenomenon mentioned in section \ref{principle}.  ``PENN" is with much fewer free parameters, but needs more training samples and training time than ``Rank" and performs worse than the other two methods in most cases (see the boldfaced values).

For interference coordination, the system performance in the table is lower, because we intend to fairly compare the training complexity
of the DNNs. Our results show that the sum rate achieved by ``Rank'' can be improved by using more samples for training, but ``W/o-Prior'' and ``PENN'' cannot. For example, the system performance of 0.99, 0.97 and 0.96 can be achieved by ``Rank'' for the cases of 10, 20 and 30 BS-user pairs with 200,000, 300,000 and 500,000 training samples, respectively, which is superior to ``W/o-Prior'' trained with doubled or even much more samples.

In Table~\ref{Hypara}, the number of neurons in hidden layers differs for different DNNs. Our further results show that the sample complexity can still be reduced significantly by exploiting the PE property when they are identical (say 300 neurons are used in the hidden layers for three DNNs in the case of 30 subcarriers or files).

\section{Concluding Remarks}
In this article, we discussed how to leverage PE property inherent in many resource allocation policies for improving the learning efficiency of DNNs. We identified several representative wireless policies that are permutation equivariant and explained why they exhibit such a property. We interpreted why PE policies can be learnt with low training complexity by using ranking, which not only simplifies the task function for a DNN to learn but also changes the input distribution.
The results in the case study showed that ranking can achieve the same system performance as the counterpart without using prior with much lower training complexity, which is more pronounced in reducing sample complexity for systems with large number of objects due to the ``sample hardening''.

Though this article considered the training of DNNs with supervision, both ranking and PENN are also applicable for the DNNs trained without labels and for other machine learning techniques. To achieve the promising gain, many issues remain open, say how to identify the objects and states of a PE policy, how to sort the samples when each object has more than one states, and how to improve the system performance of PENN and reduce the computing time for training PENN.

% \section{Acknowledgment}
% This work is supported by National Natural Science Foundation of China (NSFC) under Grant 61731002.

\bibliographystyle{IEEEtran}
\bibliography{PI-ML}

% Generated by IEEEtran.bst, version: 1.13 (2008/09/30)
\begin{thebibliography}{10}
\providecommand{\url}[1]{#1}
\csname url@samestyle\endcsname
\providecommand{\newblock}{\relax}
\providecommand{\bibinfo}[2]{#2}
\providecommand{\BIBentrySTDinterwordspacing}{\spaceskip=0pt\relax}
\providecommand{\BIBentryALTinterwordstretchfactor}{4}
\providecommand{\BIBentryALTinterwordspacing}{\spaceskip=\fontdimen2\font plus
\BIBentryALTinterwordstretchfactor\fontdimen3\font minus
  \fontdimen4\font\relax}
\providecommand{\BIBforeignlanguage}[2]{{%
\expandafter\ifx\csname l@#1\endcsname\relax
\typeout{** WARNING: IEEEtran.bst: No hyphenation pattern has been}%
\typeout{** loaded for the language `#1'. Using the pattern for}%
\typeout{** the default language instead.}%
\else
\language=\csname l@#1\endcsname
\fi
#2}}
\providecommand{\BIBdecl}{\relax}
\BIBdecl

\bibitem{LYproc2020}
L.~Liang, H.~Ye, G.~Yu, and G.~Y. Li, ``Deep-learning-based wireless resource
  allocation with application to vehicular networks,'' \emph{Proceedings of the
  IEEE}, vol. 108, no.~2, pp. 341--356, Feb. 2020.

\bibitem{RMPHY2020}
F.~Restuccia and T.~Melodia, ``Deep learning at the physical layer: System
  challenges and applications to 5{G} and beyond,'' \emph{IEEE Commun. Mag.},
  vol.~58, no.~10, pp. 58--63, Oct. 2020.

\bibitem{Spawc2017}
H.~Sun, X.~Chen, Q.~Shi, M.~Hong, X.~Fu, and N.~D. Sidiropoulos, ``Learning to
  optimize: training deep neural networks for interference management,''
  \emph{IEEE Trans. on Signal Processing}, vol.~66, no.~20, pp. 5438--5453,
  2018.

\bibitem{Samuel2019Lrn2Det}
N.~{Samuel}, T.~{Diskin}, and A.~{Wiesel}, ``Learning to detect,'' \emph{IEEE
  Trans. on Signal Processing}, vol.~67, no.~10, pp. 2554--2564, May 2019.

\bibitem{Zhou2019Beam}
P.~{Zhou}, X.~{Fang}, X.~{Wang}, Y.~{Long}, R.~{He}, and X.~{Han}, ``Deep
  learning-based beam management and interference coordination in dense mmwave
  networks,'' \emph{IEEE Trans. on Vehicular Technology}, vol.~68, no.~1, pp.
  592--603, Jan. 2019.

\bibitem{WJJ2020TCom}
J.~{Wu}, C.~{Yang}, and B.~{Chen}, ``Proactive caching and bandwidth allocation
  in heterogenous networks by learning from historical numbers of requests,''
  \emph{IEEE Trans. on Commun.}, vol.~68, no.~7, pp. 4394--4410, July 2020.

\bibitem{Alessio2019Model}
A.~{Zappone}, M.~{Di Renzo}, and M.~{Debbah}, ``Wireless networks design in the
  era of deep learning: Model-based, {AI}-based, or both?'' \emph{IEEE Trans.
  on Commun.}, vol.~67, no.~10, pp. 7331--7376, Oct. 2019.

\bibitem{Yann2015DL}
Y.~LeCun, Y.~Bengio, and G.~Hinton, ``\BIBforeignlanguage{English (US)}{Deep
  learning},'' \emph{\BIBforeignlanguage{English (US)}{Nature Cell Biology}},
  vol. 521, no. 7553, pp. 436--444, May 2015.

\bibitem{Jie2021Online}
J.~Zhang, C.~Sun, and C.~Yang, ``Resource allocation in {URLLC} with online
  learning for mobile users,'' \emph{IEEE VTC Spring}, 2021.

\bibitem{Shen2020LORM}
Y.~{Shen}, Y.~{Shi}, J.~{Zhang}, and K.~B. {Letaief}, ``{LORM}: Learning to
  optimize for resource management in wireless networks with few training
  samples,'' \emph{IEEE Trans. on Wireless Commun.}, vol.~19, no.~1, pp.
  665--679, Jan. 2020.

\bibitem{Unfolding2020}
Q.~Hu, Y.~Cai, Q.~Shi, K.~Xu, G.~Yu, and Z.~Ding, ``Iterative algorithm induced
  deep-unfolding neural networks: Precoding design for multiuser {MIMO}
  systems,'' \emph{IEEE Trans. on Wireless Commun.}, p. early access, 2020.

\bibitem{Zaheer2017DeepSets}
Z.~Manzil, K.~Satwik, R.~Siamak, P.~Barnabas, S.~Ruslan, and S.~Alexander,
  ``Deep sets,'' \emph{Advances in Neural Information Processing Systems},
  2017.

\bibitem{GJ2020ICC}
J.~Guo and C.~Yang, ``Structure of deep neural networks with a priori
  information in wireless tasks,'' \emph{IEEE ICC}, 2020.

\bibitem{Eisen2020}
M.~Eisen and A.~Ribeiro, ``Optimal wireless resource allocation with random
  edge graph neural networks,'' \emph{IEEE Trans. on Signal Processing},
  vol.~68, no.~10, pp. 2977--2991, 2020.

\bibitem{OSbook}
H.~A. David and H.~N. Nagaraja, \emph{Order Statistics}.\hskip 1em plus 0.5em
  minus 0.4em\relax John Wiley and Sons, 2003.

\end{thebibliography}

% \section{Biography}
% Chengjian Sun is a Ph.D. student of Beihang University. His research interests include machine learning for optimization and URLLC.

% Jiajun Wu is a Ph.D. student of Beihang University. His research interests include edge caching and machine learning for wireless systems.

% Chenyang Yang is a professor of Beihang University. She has served as associate or guest editors for several IEEE journals and TPC members, TPC co-chair or Track co-chair for many IEEE conferences. Her current research interests lie in edge intelligence for 5G and beyond.

\end{document}